# Sensitivity analysis of a mean-value exergy-based internal combustion engine model

Gabriele Pozzato*, Denise M. Rizzo**, and Simona Onori*
* Energy Resources Engineering, Stanford University, ** Ground Vehicle Systems Center, U.S. Army CCDC

## Abstract

In this work, we conduct a sensitivity analysis of the mean-value internal combustion engine exergy-based model, recently developed by the authors, with respect to different driving cycles, ambient temperatures, and exhaust gas recirculation rates. Such an analysis allows to assess how driving conditions and environment affect the exergetic behavior of the engine, providing insights on the system's inefficiency. Specifically, the work is carried out for a military series hybrid electric vehicle.

## Introduction

To improve the energy efficiency and reduce operational costs, military ground vehicles are migrating from conventional internal combustion engine (ICE) powertrains to electrified ones. To unleash the full potential of these technologies, mathematical modeling is a fundamental tool to assess and quantify the system's irreversibilities (responsible for efficiency degradation) [1]. Availability (or exergy) is used to measure the mechanical, chemical, and thermal work a system can deliver with respect to a reference state. On the contrary to approaches based on the first law of thermodynamics, techniques based on exergy principles rely on the second law of thermodynamics. This allows to locate the system's irreversibilities and develop strategies for efficiency improvement.

In the aerospace field, exergy principles have been widely used for design optimization and system-level analysis. Some examples are the design optimization of hypersonic aircrafts [2,3] and the exergetic analysis of launch vehicles [4] and rockets [5]. Availability modeling has been applied to ICEs with the goal of quantifying irreversibilities and, ultimately, optimize combustion. The authors of [6] and [7] analyze the exergy transfer and destruction phenomena in spark ignition engines. In [8] and [9], compression ignition engines operating in steady-state and transient conditions are investigated. The effect of different types of fuel (biodiesel and bioethanol) to the availability balance is analyzed in [10] and [11]. Other applications of exergy principles can be found in power generation [12,13] and naval engineering [14].

In [15], the authors develop an exergy-based modeling framework for electric and hybrid electric vehicles (EVs and HEVs). The proposed framework allows the assessment of availability transfer and destruction phenomena in the powertrain. In HEVs, the ICE is the principal source of inefficiency and must be carefully analyzed. To this aim, the authors have recently proposed an innovative mean-value exergy-based model for ICEs [16]. This model accounts for all the availability transfer and destruction phenomena. A novel characterization of combustion irreversibilities, based on the quantification of entropy generation in chemical reactions, is formulated. Moreover, a model of the thermal exchange between the in-cylinder mixture and the cylinder wall based on the Hohenberg correlation is used. This ICE exergy model provides useful information on the engine efficiency for each operating point in the speed-torque plane while keeping the computational burden low.

In this paper, the ICE exergy model developed in [16] is applied to a military series HEV case study and a thorough sensitivity analysis with respect to different driving conditions is performed. The exergy balance is defined with respect to the environment, which is characterized by a certain pressure, chemical composition, and temperature. Composition and pressure are assumed to remain constant, and modifications of the environmental temperature are carefully analyzed. In diesel engines, exhaust gas recirculation (EGR) is used to mitigate hazardous $NO_x$ emissions. Varying the EGR rate leads to modifications of the mixture composition and of the combustion process. This affects the exergy term related to combustion irreversibilities (or, in other words, the ICE inefficiency) and must be investigated carefully. Starting from four military driving cycles, the model is simulated for different environmental temperature and EGR rate combinations. Correlations between the exergy transfer/destruction phenomena and modifications of environmental temperature and EGR rate are investigated. Ultimately, the mean-value ICE exergy model allows to compute the engine availability balance in diverse driving scenarios, providing insights on how driving conditions and environment affect the exergetic behavior of the engine.

The paper is organized as follows. First, the exergy-related concepts from thermodynamics are summarized [15]. Then, a detailed ICE exergy-based model is described with focus on its mean-value implementation. Thirdly, the exergy balance for a constant power request is shown. The sensitivity analysis is conducted for a Power Stroke 6.4L V8 diesel engine and results considering different driving cycles, ambient temperatures, and EGR rates are shown. Finally, conclusions are carried out.

Table 1. ICE parameters [16].

| Parameter | Value | Unit |
|---|---|---|
| $f_{N_2,0}$ | 0.7567 | [-] |
| $f_{CO_2,0}$ | 0.0003 | [-] |
| $f_{H_2O,0}$ | 0.0303 | [-] |
| $f_{O_2,0}$ | 0.2035 | [-] |
| $f_{others,0}$ | 0.0092 | [-] |
| $r_c$ | 17.5:1 | [-] |
| $n_{cyl}$ | 8 | [-] |
| $x_{EGR}$ | 0.2 | [-] |
| $x$ | 14.4 | [-] |
| $y$ | 24.9 | [-] |
| $V_{d,tot}$ | 6.4 | [l] |
| $T_0$ | 293.15 | [K] |
| $T_I$ | 323.15 | [K] |
| $P_0$ | 1 | [bar] |
| $P_I$ | 1 | [bar] |
| $C_1$ | 75 | [kPa] |
| $C_2$ | 0.458 | [s kPa] |
| $C_3$ | 0.4 | [$s^2 kPa/m^2$] |
| $R_{gas}$ | 8.31 | [J/(mol K)] |
| LHV | 42.50 | [MJ/kg] |

## Exergy modeling: definitions

Theoretical concepts on exergy analysis and modeling are summarized. For further information on exergy principles, the reader is referred to [15].

*Definition 2.1 (Exergy).* Useful work that can be extracted from a system at a given state with respect to a chemical and thermodynamic reference state.

*Definition 2.2 (Physical exergy).* Work potential between the current and restricted state of the system.

*Definition 2.3 (Chemical exergy).* Work potential associated with the different chemical composition of the reference and restricted state.

*Definition 2.4 (Reference state).* The reference state is characterized by a temperature $T_0$, a pressure $P_0$, and a chemical composition defined by molar fractions $f_0$. At the reference state, the available work (mechanical, thermodynamical, chemical, etc.) is zero.

*Definition 2.5 (Restricted state).* State of a system not in chemical equilibrium with the reference state.

Quantities expressed with respect to the reference and restricted state are denoted by subscript 0 and superscript $\star$, respectively.

## Internal combustion engine exergy model

### Engine technical specifications

The ICE model is developed considering a Power Stroke 6.4L V8 diesel engine. The engine is equipped with a turbocharger and an EGR system and is characterized by a peak power of 260kW at 3000rpm. The properties of the ICE are listed in Table 1.

Figure 1 shows the fuel rate $\dot{m}_f$ (a) and exhaust gas temperature $T_E$ (b) as a function of the engine operating point.

### In cylinder mean-value model

For an effective characterization of the availability transfer and destruction phenomena, a careful description of the in-cylinder combustion dynamics is necessary. The model of the internal combustion engine is developed under the following assumptions [16]:
- during combustion, all the fuel injected in the cylinder is burnt;
- the engine operates in steady-state conditions, i.e., torque actuation and exhaust transport delays are not taken into account;
- all cylinders behave identically, i.e., given an operating point there is no difference in the combustion process between cylinders.

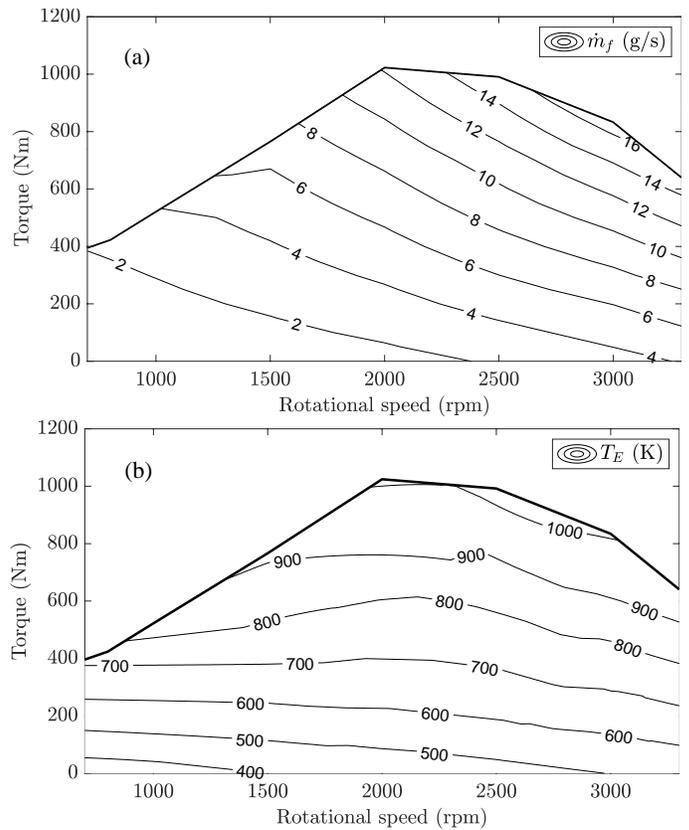

Figure 1. (a) Fuel rate ($\dot{m}_f$) and (b) exhaust gas temperature ($T_E$) maps for a Power Stroke 6.4L V8 diesel engine.

First, the in-cylinder crank-angle resolved model is formulated relying on a single-zone approach. Working fluids are modeled as one thermodynamic system experiencing mass and energy transfer with the surroundings. Spatial homogeneity of temperature and pressure is assumed. Moreover, burnt and unburnt fractions of the mixture are not separated, and the heat released is assumed to be evenly distributed. Starting from in-cylinder crank-angle resolved quantities (namely, piston motion, pressure, fuel burning dynamics, and thermal exchange dynamics), the mean-value in-cylinder pressure, temperature, and heat transfer are derived averaging the corresponding crank-angle based quantities over the firing interval. The computation of the mean values over the firing interval captures the heat transferred and released during the combustion process around the top dead center (TDC) (where most of the heat is released).

Figures 2, 3, and 4 show the in-cylinder mean-value pressure ($P_{cyl}$), temperature ($T_{cyl}$), and heat transfer ($\dot{Q}_{cyl}$) for each engine operating point. These maps are obtained considering 0.2 EGR rate and $T_0 = 293.15K$ (Table 1). An increase of the load is related to higher fueling levels (Figure 1a) and leads to higher in-cylinder pressure, temperature, and heat transfer.

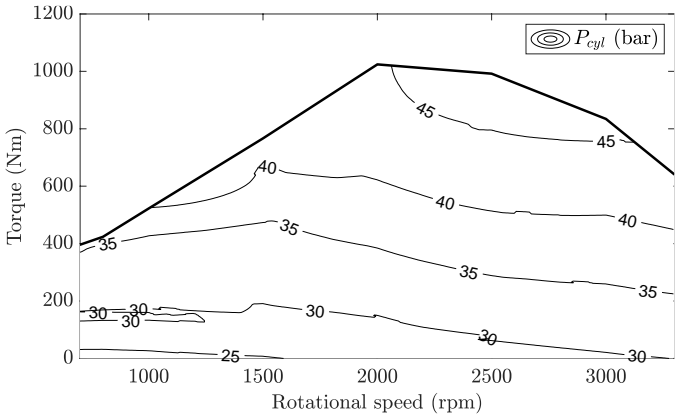

Figure 2. In-cylinder average pressure. The EGR rate is set to 0.2 and $T_0 = 293.15K$.

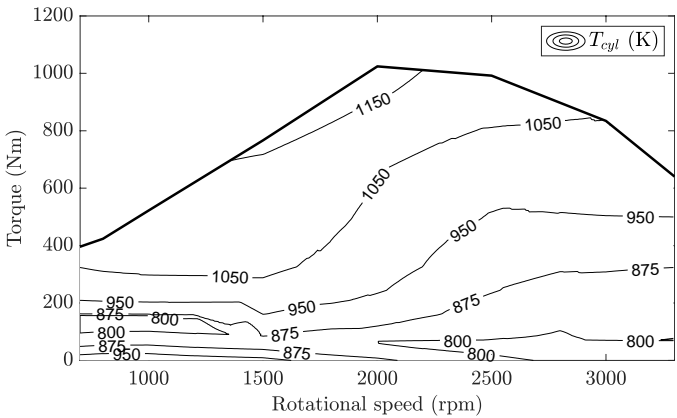

Figure 3. In-cylinder average temperature. The EGR rate is set to 0.2 and $T_0 = 293.15K$.

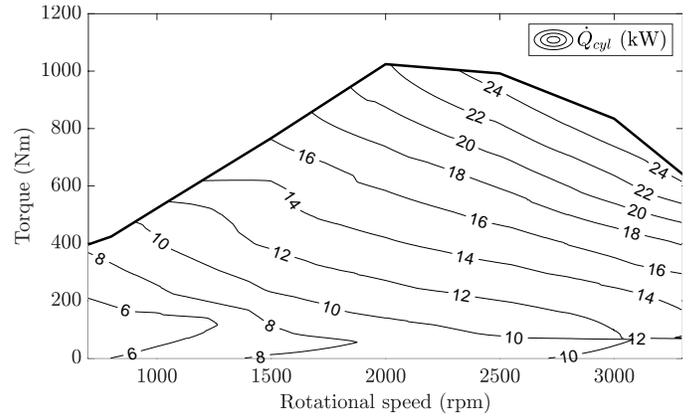

Figure 4. Engine average heat transfer between in-cylinder mixture and walls. The EGR rate is set to 0.2 and $T_0 = 293.15K$.

*Exergy mean-value model*

Relying on maps in Figures 2, 3, and 4, the exergetic behavior of the ICE is analyzed considering the mean-value model developed in [16].

The exergy balance is defined as follows:
$$\dot{X}_{eng} = \dot{X}_{In} + \dot{X}_{Out} + \dot{X}_{Dest} + \dot{X}_{others} =$$
$$= \dot{X}_{fuel,eng} + \dot{X}_{intk,eng} + \dot{X}_{work,eng} + \dot{X}_{heat,eng} + \quad (1)$$
$$+ \dot{X}_{exh,eng} + \dot{X}_{comb,eng} + \dot{X}_{fric,eng} + \dot{X}_{others}$$

with $\dot{X}$ the ICE exergy transfer and destruction phenomena. The mathematical formulation of each term is shown in Table 2.

Fuel and intake air are the exergy flows entering the control volume: $\dot{X}_{In} = \dot{X}_{fuel,eng} + \dot{X}_{intk,eng}$. The fuel availability $\dot{X}_{fuel,eng}$ is the capability of doing work associated with the chemical exergy stored in the injected fuel and is modeled in equation (T.1). The intake exergy flow, equation (T.2), considers both chemical and physical exergies of the mixture entering the cylinder. Given the presence of EGR, the mixture is a combination of fresh air and recirculated exhaust gas.

The mechanical work ($\dot{X}_{work,eng}$), heat exchange ($\dot{X}_{heat,eng}$), and exhaust transport ($\dot{X}_{exh,eng}$) terms transfer the exergy out of the control volume ($\dot{X}_{Out}$). These terms are defined by equations (T.3), (T.4), and (T.5), respectively.

In the ICE, exergy is destroyed by combustion irreversibilities ($\dot{X}_{comb,eng}$) and frictions ($\dot{X}_{fric,eng}$). The former is related to the entropy generated during combustion and is function of the in-cylinder temperature, i.e., the higher the temperature, the higher the efficiency, and the lower the entropy generation and combustion irreversibilities. The latter models friction losses due to components motion and turbulent dissipation. Together, these terms (modeled as in equations (T.6) and (T.7)) constitute $\dot{X}_{Dest}$, i.e., the destroyed availability.

$\dot{X}_{others}$ (equation (T.8)) lumps all the unmodeled exergy destruction and transfer phenomena: losses in valves throttling, nonuniform in-cylinder combustion, and blow-by gases. $\dot{X}_{others}$ is computed inverting equation (1), under the assumption that the engine is working in steady-state conditions (transient phenomena between operating points are neglected). According to [16] and [17], this condition is rewritten as:
$$\dot{X}_{eng} = 0 \qquad (2)$$

## Table 2. ICE mean-value exergy model.

**Fuel** (T.1)

$$\dot{X}_{fuel,eng} = \left(1.04224 + 0.011925 \frac{x}{y} - \frac{0.042}{x}\right) LHV \, \dot{m}_f$$

**Intake gas** (T.2)

$$\dot{X}_{intk,eng} = \sum_{\sigma \in \mathcal{S}} \dot{n}_{I,\sigma} \left(\psi^I_{ch,\sigma} + \psi^I_{ph,\sigma}\right), \qquad \mathcal{S} = \{N_2, CO_2, H_2O, O_2\}$$

$$\dot{n}_{I,\sigma} = \dot{n}_I \, f^I_\sigma$$

$$\psi^I_{ph,\sigma} = \left(h_\sigma(T_I) - T_0 s_\sigma(T_I)\right) - \left(h^\star_\sigma(T_0) - T_0 s^\star_\sigma(T_0)\right)$$

$$\psi^I_{ch,\sigma} = R_{gas} T_0 \log\left(\frac{f^I_\sigma}{f_{\sigma,0}}\right)$$

**Mechanical work** (T.3)

$$\dot{X}_{work,eng} = -T_{eng} \omega_{eng} = -P_{eng}$$

**Heat exchange** (T.4)

$$\dot{X}_{heat,eng} = \left(1 - \frac{T_0}{T_{cyl}}\right)(-\dot{Q}_{cyl})$$

**Exhaust gas** (T.5)

$$\dot{X}_{exh,eng} = -\sum_{\sigma \in \mathcal{S}} \dot{n}^E_{E,\sigma} \left(\psi^E_{ch,\sigma} + \psi^E_{ph,\sigma}\right)$$

$$\dot{n}_{E,\sigma} = \dot{n}_E \, f^E_\sigma$$

$$\psi^E_{ph,\sigma} = \left(h_\sigma(T_E) - T_0 s_\sigma(T_E)\right) - \left(h^\star_\sigma(T_0) - T_0 s^\star_\sigma(T_0)\right)$$

$$\psi^E_{ch,\sigma} = R_{gas} T_0 \log\left(\frac{f^E_\sigma}{f_{\sigma,0}}\right)$$

**Combustion irreversibilities** (T.6)

$$\dot{X}_{comb,eng} = -\frac{T_0}{T_{cyl}}\left\{g_f - x \, g_{CO_2} - \frac{y}{2} g_{H_2O} + \left(x + \frac{y}{4}\right) g_{O_2}\right\} \dot{n}_f +$$

$$-\frac{T_0}{T_{cyl}}\left\{\frac{\lambda}{1 - x_{EGR}}\left(x + \frac{y}{4}\right) 3.76 R_{gas} T_{cyl} \log\left(\frac{f^I_{N_2}}{f^E_{N_2}}\right)\right\} \dot{n}_f +$$

$$-\frac{T_0}{T_{cyl}}\left\{R_{gas} T_{cyl} \sum_{\sigma \in \mathcal{S} \setminus \{N_2\}} \left[\nu^I_\sigma \log\left(\frac{f^I_\sigma P_{cyl}}{P_0}\right) - \nu^E_\sigma \log\left(\frac{f^E_\sigma P_{cyl}}{P_0}\right)\right]\right\} \dot{n}_f$$

**Frictions** (T.7)

$$\dot{X}_{fric,eng} = -\frac{\omega_{eng}}{4\pi} FMEP \, V_{d,tot}$$

$$FMEP = 1000[C_1 + C_2 \omega_{eng} + C_3 S_p^2]$$

**Others** (T.8)

$$\dot{X}_{others} = -\left[\dot{X}_{fuel,eng} + \dot{X}_{intk,eng} + \dot{X}_{work,eng} + \dot{X}_{heat,eng}\right] +$$
$$-\left[\dot{X}_{exh,eng} + \dot{X}_{comb,eng} + \dot{X}_{fric,eng}\right]$$

Terms in Table 2 are exergy rates, i.e., they express the variation of exergy with respect to time. Starting from these quantities, the exergy is obtained integrating over time the corresponding rate term:

$$X_k = \int_0^T \dot{X}_k \, dt \qquad (3)$$

with $\dot{X}_k \in \{\dot{X}_{fuel,eng}, \dot{X}_{intk,eng}, \dot{X}_{work,eng}, \dot{X}_{heat,eng}, \dot{X}_{exh,eng}, \dot{X}_{comb,eng}, \dot{X}_{fric,eng}, \dot{X}_{others}\}$ and $T$ the time horizon.

## Exergy balance for a constant power request

As an example, the exergy balance is computed for a constant engine power of 106kW, actuated at $\omega_{eng} = 1973$rpm and $T_{eng} = 512$Nm (average speed-torque values given the ICE maps in Figure 1). According to Table 1, the reference temperature $T_0$ is equal to 293.15K and the EGR rate is set to 0.2 (the static maps in Figures 2, 3, and 4 are used).

The percentage contribution of each availability term is computed as:

$$X_j \, [\%] = \frac{X_j}{X_{fuel,eng} + X_{intk,eng}} \times 100 \qquad (4)$$

with $X_j \in \{X_{work,eng}, X_{heat,eng}, X_{exh,eng}, X_{comb,eng}, X_{fric,eng}, X_{others}\}$. The denominator is the exergy introduced in the system ($X_{In}$), which can be transferred or destroyed according to $X_j$. Given a constant operating point, the exergy rate terms are constant and the percentage contribution in equation (4) does not change with the time horizon $T$.

The percentage contribution of the exergy transfer and destruction phenomena is:

$$\begin{aligned} X_{exh,eng} \, [\%] &= 17.7\%, & X_{comb,eng} \, [\%] &= 31.7\% \\ X_{fric,eng} \, [\%] &= 6.92\%, & X_{work,eng} \, [\%] &= 36.8\% \\ X_{heat,eng} \, [\%] &= 3.74\%, & X_{others} \, [\%] &= 3.14\% \end{aligned} \qquad (5)$$

Results are comparable to ranges shown in [11] for diesel engines, i.e., the indicated work ($X_{work,eng} + X_{fric,eng}$) accounts for 40-45%, the heat transfer term at most for 10%, the exhaust gas for 10-20%, and the combustion irreversibilities for ~25%. In this scenario, combustion irreversibilities are higher than 25%, however, values in [11] are indicative and function of the engine specifications and operating conditions.

## Sensitivity analysis

The sensitivity analysis is carried out considering the military series HEV in Figure 5. The whole model of the HEV, described in [18], is simulated considering the driving cycles in Figure 6. The solid speed profile acts as a reference which can or cannot be fulfilled depending on the requested acceleration and velocity (the actuated speed profiles are shown in Figure 6, dashed lines). The energy management strategy is based on the Pontryagin minimum principle (PMP) [21], [19] and optimizes the power split between the battery pack and ICE, minimizing a cost function accounting for fuel consumption while ensuring charge sustaining conditions. In this work, the effect of different driving conditions on the ICE exergy balance (modeled as in Table 2) is analyzed. The analysis of the exergetic behavior of other components of the powertrain (such as battery pack and electric motors) is out of the scope of this work.

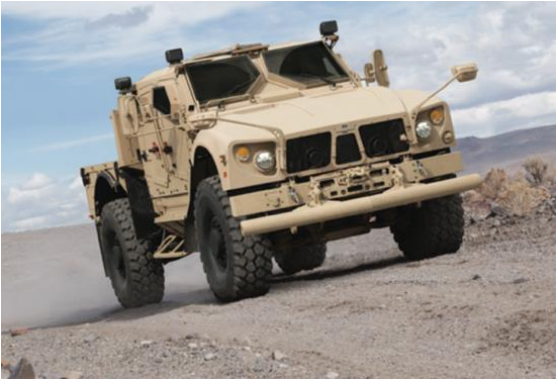

Figure 5. Oshkosh M-ATV [20].

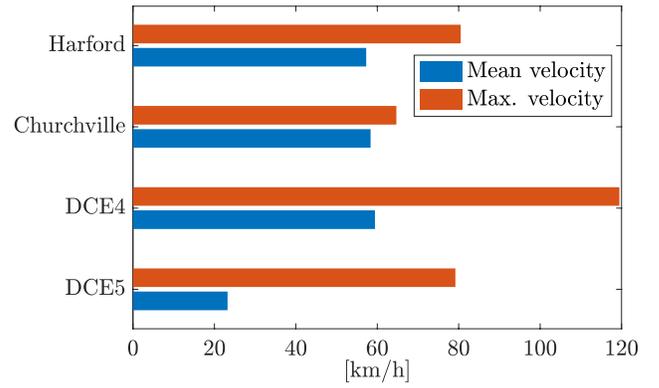

Figure 7. Mean and maximum velocities for speed profiles in Figure 6.

Driving cycles in Figure 6 provide a good combination of high- and low-speed scenarios. Figure 7 shows that DCE4 reaches the highest speed (~120km/h) and DCE5 is characterized by the lowest mean velocity. For each driving cycle, the following combinations of reference temperature ($T_0$) and EGR rate ($x_{EGR}$) are analyzed:

$$T_0 \in \{263.15, 273.15, 283.15, 293.15, 303.15, 313.15\} \text{ K} \quad (6)$$
$$x_{EGR} \in \{0, 0.1, 0.2, 0.3\}$$

Changing $T_0$ corresponds to modifying the environmental condition in which the vehicle is operating over a mission. In this work, temperatures between $-10°C$ and $40°C$ are considered. Being the reference state, $T_0$ is always the lowest temperature in the system. Instead, a modification of the EGR rate leads to a variation of the chemical composition of the in-cylinder mixture. This condition is interesting because different EGR rate could be used for emissions mitigation. A variation of the EGR rate modifies the combustion dynamics. Therefore, maps in Figures 2, 3, and 4 must be computed for each value of $x_{EGR}$, relying on the model summarized in Section "Internal combustion engine exergy model". According to [17], EGR values in equation (6) are reasonable for diesel engines.

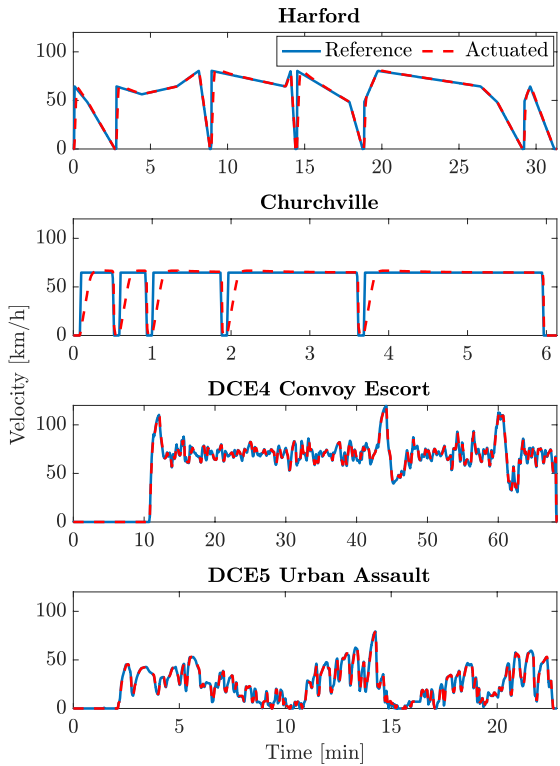

Figure 6. Driving cycles used in the sensitivity analysis. Reference (solid lines) and actuated (dashed lines) velocities are shown.

## Results

Figures A1, A2, A3, and A4 (collected in the Appendix) show the percentage contribution of the different exergy transfer and destruction phenomena for each of the driving cycle in Figure 6 and combinations of $T_0$ and $x_{EGR}$ in (6).

Results for the Harford driving cycle (Figure A1) are analyzed. The increase of the EGR rate leads to an increment of the recirculated exhaust gas mass and increases the in-cylinder thermal inertia. This decreases the in-cylinder temperature $T_{cyl}$, the heat transfer $\dot{Q}_{cyl}$, and, consequently, the exergy term $X_{heat,eng}$ (Figure A1c). Combustion reactions become less efficient at lower in-cylinder temperatures $T_{cyl}$. Therefore, increasing $x_{EGR}$ leads to increased entropy generation and to a higher contribution of $X_{comb,eng}$ to the total exergy balance (Figure A1e). The additional mass introduced in the engine by the EGR is transferred outside the system, with an increase of $X_{exh,eng}$ (Figure A1d). For practical purposes, the contribution of the indicated work ($X_{work,eng} + X_{fric,eng}$) to the exergy balance can be considered constant while varying the EGR rate (Figure A1a and A1b). The slight reduction of $X_{work,eng}$ and $X_{fric,eng}$ – when increasing $x_{EGR}$ – is caused by the intake exergy flux ($X_{intk,eng}$) increase due to the additional mass of recirculated exhaust gas. $X_{others}$ remains constrained to values in line with the literature [11] and increases moving towards $x_{EGR} = 0$ (Figure A1f). However, this condition is unlikely for diesel engines because it would lead to unacceptable $NO_x$ emissions.

An increase of the reference temperature $T_0$ leads to a reduction of $X_{exh,eng}$ (Figure A1d) caused by a lower physical exergy $\psi_{ph}$, which translates to a reduced capability of the exhaust gas of doing work (the effect of the chemical exergy $\psi_{ch}$ is negligible, i.e., it is always one order of magnitude smaller than $\psi_{ph}$). The increase of $T_0$ leads to a reduction of the Carnot efficiency (the temperature difference between $T_0$ and $T_{cyl}$ in equation (T.4) is reduced) and to a lower $X_{heat,eng}$ (Figure A1c). Moreover, increasing $T_0$ has a similar effect of reducing $T_{cyl}$ and leads to an increased $X_{comb,eng}$ (Figure A1e). As for the EGR case, the contribution of the indicated work ($X_{work,eng} + X_{fric,eng}$) to the exergy balance can be considered constant while varying $T_0$ (Figure A1a and A1b). The slight increase of $X_{work,eng}$ and $X_{fric,eng}$ while increasing $T_0$ is caused by the intake exergy flux decrease ($\psi_{ch} + \psi_{ph}$ in equation (T.2)). $X_{others}$ remains constrained to values in line with the literature and decreases while increasing $T_0$. The effect of $T_0$ on $X_{others}$ is lower if compared to varying $x_{EGR}$ (Figure A1f).

In Figures A2, A3, and A4, results for Churchville, DCE4 Convoy Escort, and DCE5 Urban Assault cycles are shown. Trends described for Figure A1 apply also to these driving conditions. Figure 8 compares simulation results at $x_{EGR} = 0.2$ and $T_0 = 293.15K$ (nominal conditions defined in Table 1) for each driving cycle. Computing the standard deviation (SD) shows that the percentage contribution of each exergy transfer and destruction term does not change drastically while modifying the driving style: SDs are always below 2%. This is due to the PMP-based energy management strategy that uses the ICE only in operating points characterized by the highest efficiency, avoiding exploring the full speed-torque plane.

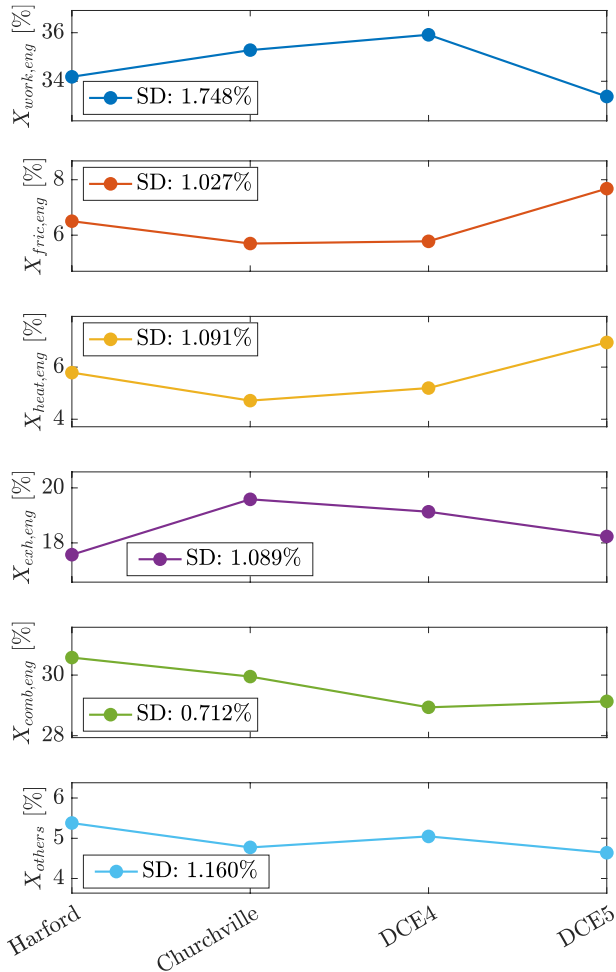

Figure 8. Percentage contribution of the exergy transfer and destruction terms for each driving cycle at $x_{EGR} = 0.2$ and $T_0 = 293.15K$.

## Summary/Conclusions

This paper shows the application of the mean-value ICE exergy model developed in [16] to a military series HEV. A comprehensive sensitivity analysis with respect to different driving cycles, environmental temperatures, and EGR rates is carried out. The analysis shows that an increase of $T_0$ leads to a decrease of the heat exchange and exhaust gas exergy terms and to an increase of the mechanical work, frictions, and combustion irreversibilities components. Increasing $x_{EGR}$ leads to a decrease of the mechanical work, frictions, and heat exchange terms and to an increment of combustion irreversibilities and exhaust gas availability.

The analysis proves the effectiveness of the mean-value ICE exergy model for the quantification of availability transfer and destruction phenomena in different driving conditions. This enables the use of such a model inside the framework proposed in [15], providing a useful tool for the future development of management strategies aimed at minimizing ground vehicle exergy losses.

## Contact Information

Gabriele Pozzato, Energy Resources Engineering, Stanford University, Stanford, CA 94305 (e-mail: gpozzato@stanford.edu).

## Acknowledgments

Unclassified. DISTRIBUTION STATEMENT A. Approved for public release; distribution is unlimited. Reference herein to any specific commercial company, product, process, or service by trade name, trademark, manufacturer, or otherwise does not necessarily constitute or imply its endorsement, recommendation, or favoring by the United States Government or the Dept. of the Army (DoA). The opinions of the authors expressed herein do not necessarily state or reflect those of the United States Government or the DoD, and shall not be used for advertising or product endorsement purposes.

## Definitions/Abbreviations

| Symbol | Description |
| --- | --- |
| $\lambda$ | Air-fuel equivalence ratio [-] |
| $f$ | Mole fraction [-] |
| $n_{cyl}$ | Number of cylinders [-] |
| $r_c$ | Compression ratio [-] |
| $x, y$ | Fuel chemical formula coefficient [-] |
| $x_{EGR}$ | EGR rate [-] |
| $t, dt$ | Time and its differential [s] |
| $T$ | Time horizon [s] |
| $V_{d,tot}$ | Engine displacement [m$^3$] |
| $S_p$ | Mean piston speed [m/s] |
| $\nu$ | Stoichiometric coefficient [mol] |
| $n, \dot{n}$ | Moles and molar flow rate [mol], [mol/s] |
| $m, \dot{m}$ | Mass and mass flow rate [kg], [kg/s] |
| $FMEP$ | Friction mean effective pressure [Pa] |
| $C_1, C_2, C_3$ | FMEP coefficients [kPa], [s kPa], [s$^2$kPa] |
| $P_0$ | Reference state pressure [bar] |
| $P_I$ | Intake gas pressure [bar] |
| $P_{cyl}$ | In-cylinder gas pressure [bar] |
| $T_0$ | Reference state temperature [K] |
| $T_I, T_E$ | Intake and exhaust gas temperature [K] |
| $T_{cyl}$ | In-cylinder gas temperature [K] |
| $\psi_{ch}, \psi_{ph}$ | Chemical and physical exergy flux [J/mol] |
| $g$ | Gibbs free energy [J/mol] |
| $h$ | Specific enthalpy [J/mol] |
| $R_{gas}$ | Ideal gas constant [J/(mol K)] |
| $s$ | Specific entropy [J/(mol K)] |
| $LHV$ | Fuel lower heating value [MJ/kg] |
| $X, \dot{X}$ | Exergy and exergy rate [J],[W] |
| $\dot{Q}_{cyl}$ | Heat release during combustion [W] |
| $P_{eng}$ | ICE power [W] |
| $\omega_{eng}$ | ICE rotational speed [rad/s] |
| $T_{eng}$ | ICE torque [Nm] |
| 0 | Reference state |
| $\star$ | Restricted state |
| $\sigma$ | Chemical species |
| $\mathcal{S}$ | Set collecting the chemical species $\sigma$, defined as $\{N_2, CO_2, H_2O, O_2\}$ |
| I,E | Intake and exhaust |
| EGR | Exhaust gas recirculation |

| EV | Electric vehicle | PMP | Pontryagin minimum principle |
| HEV | Hybrid electric vehicle | SD | Standard deviation |
| ICE | Internal combustion engine | TDC | Top dead center |

# Appendix

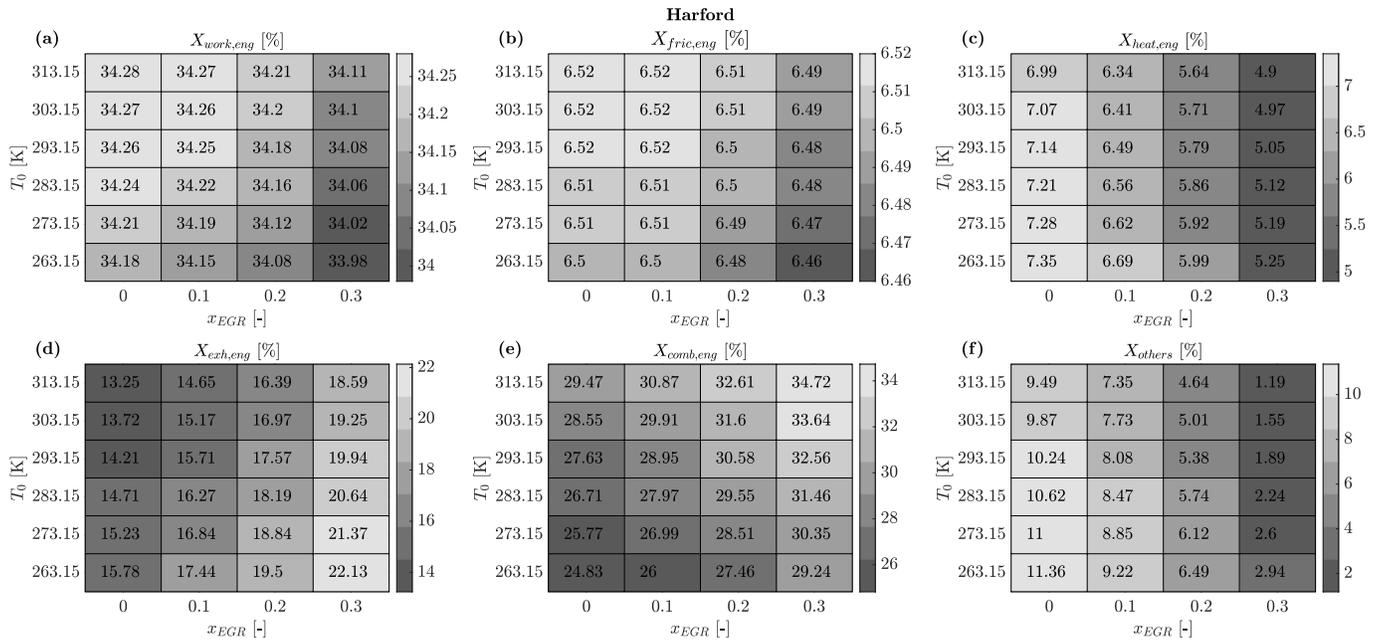

Figure A1. Harford – Percentage contribution of the exergy transfer and destruction terms as a function of the EGR rate ($x_{EGR}$) and reference temperature ($T_0$). The percentage contribution is computed with respect to the input exergy $X_{fuel,eng} + X_{intk,eng}$.

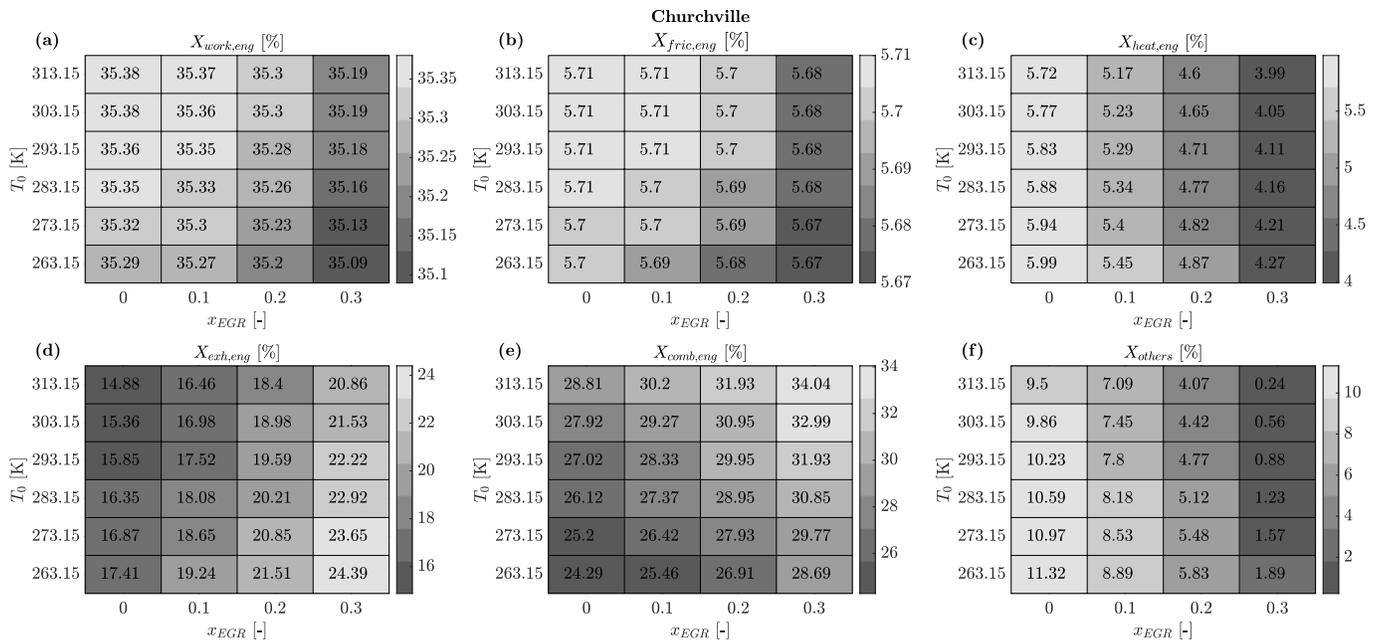

Figure A2. Churchville – Percentage contribution of the exergy transfer and destruction terms as a function of the EGR rate ($x_{EGR}$) and reference temperature ($T_0$). The percentage contribution is computed with respect to the input exergy $X_{fuel,eng} + X_{intk,eng}$.

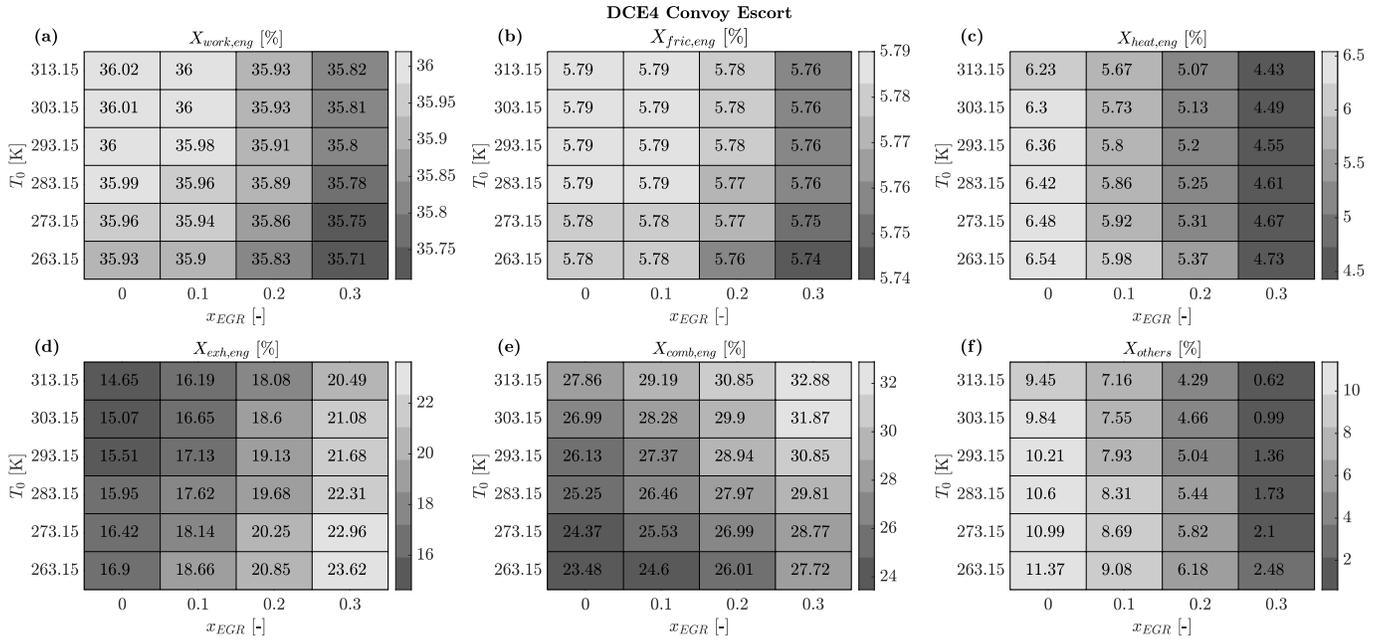

Figure A3. DCE4 Convoy Escort − Percentage contribution of the exergy transfer and destruction terms as a function of the EGR rate ($x_{EGR}$) and reference temperature ($T_0$). The percentage contribution is computed with respect to the input exergy $X_{fuel,eng} + X_{intk,eng}$.

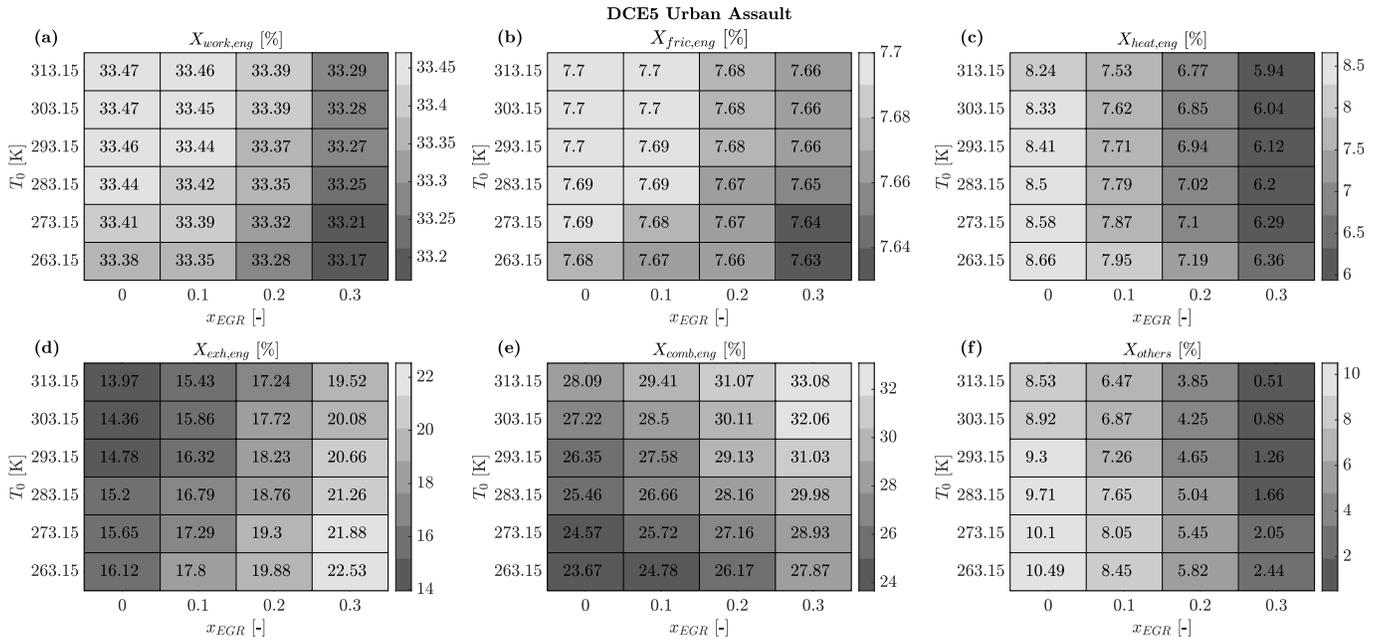

Figure A4. DCE5 Urban Assault − Percentage contribution of the exergy transfer and destruction terms as a function of the EGR rate ($x_{EGR}$) and reference temperature ($T_0$). The percentage contribution is computed with respect to the input exergy $X_{fuel,eng} + X_{intk,eng}$.